\documentclass[preprint]{elsarticle}

\usepackage{booktabs} 
\usepackage{subcaption}
\usepackage{multirow}
\usepackage{amsmath}
\usepackage{fancyvrb}
\usepackage{color}
\usepackage{amsthm}
\usepackage{amsmath}

\usepackage[margin=1in]{geometry}

\usepackage{paralist, tabularx}

\usepackage{hyperref}

\usepackage{multirow}
\usepackage[numbers]{natbib}
\usepackage{algorithm}
\usepackage{algorithmic}

\usepackage{makecell} 

\journal{Journal of Information Security and Applications}

\begin{document}

\begin{frontmatter}

\title{On the Construction of a Post-Quantum Blockchain for Smart City}

\tnotetext[t1]{\textcolor{blue}{{\bf This work has been accepted by Journal of Information Security and Applications, Elsevier, 2021.}}}


\author[1]{Jiahui Chen}
\ead{csjhchen@gmail.com}

\author[2]{Wensheng Gan\corref{ca}} 
\ead{wsgan001@gmail.com}
\cortext[ca]{Corresponding author}

\author[3]{Muchuang Hu}
\ead{moonchallenge@gmail.com}

\author[4]{Chien-Ming Chen}
\ead{chienmingchen@ieee.org}

\address[1]{School of Computer, Guangdong University of Technology, Guangzhou 510006, China}
\address[2]{College of Cyber Security, Jinan University, Guangzhou 510632, China}
\address[3]{People's Bank of China Guangzhou, Guangzhou 510006, China}
\address[4]{College of Computer Science and Engineering, Shandong University of Science and Technology, Qingdao 266590, China}

\begin{abstract}

Owing to some special characteristics and features, blockchain is a very useful technique that can securely organize diverse devices in a smart city. It finds wide applications, especially in distributed environments, where entities such as wireless sensors need to be certain of the authenticity of the server. As contemporary blockchain techniques that address post-quantum concerns have not been designed, in this study, we investigate a blockchain in the post-quantum setting and seek to discover how it can resist attacks from quantum computing. In addition, traditional proof of work (PoW)-based consensus protocols such as Bitcoin cannot supply memory mining, and the transaction capacity of each block in a blockchain is limited and needs to be expanded. Thus, a new post-quantum proof of work (post-quantum PoW) consensus algorithm for security and privacy of smart city applications is proposed. It can be used to not only protect a blockchain under a quantum computing attack compared to  existing classical hash-based PoW algorithms but also to supply memory mining. Meanwhile, an identity-based post-quantum signature is embedded into a transaction process to construct lightweight transactions. Subsequently, we provide a detailed description on the execution of the post-quantum lightweight transaction in a blockchain. Overall, this work can help enrich the research on future post-quantum blockchain and support the construction or architecture of emerging blockchain-based smart cities.

\end{abstract}


\end{frontmatter}


\section{Introduction}

The characteristic that some specified devices can be securely organized in a distributed way is very attractive in smart cities \cite{xie2019survey}. In many previous studies, blockchains have been shown to be very helpful, especially for applications in smart city environments where entities such as wireless sensors need to be certain of the authenticity of the server \cite{xie2019survey}. Blockchain technology was developed with the invention of Bitcoin in 2008 by Nakamoto \cite{Nakamoto2008}, which was considered as the backbone of cryptocurrency, and, in recent years, its success has attracted much organizational research on the use of blockchain technology. Along with the increasing price of Bitcoins, which is currently more than 10000 dollars, and popular projects such as Ethereum \cite{Wood2014}, Monero \cite{Van2013}, and Ripple \cite{Armknecht2015}, decentralized blockchain has attracted people's attention and can be applied to various trades.

It raises several security and privacy issues in blockchain-based smart cities. To summarize, there are two main parts in the cryptocurrency application of the modern blockchain: a consensus protocol that is used to create new blocks and a cryptographic signature scheme that is used to verify transactions. Details are presented below.

First, a blockchain is essentially a distributed consensus storage system, with consensus protocols between nodes to agree on the contents of the storage, ensuring that the ledger stored by each node in the distributed network is consistent. It can reach consensus among different parties in an open, untrusted network with security guaranteed by a carefully designed consensus mechanism along with crypto mechanisms such as hash and public key cryptography. Therefore, consensus protocols are one of the key technologies in blockchain ecosystems. In fact, with the development of blockchains, a number of projects depend on different consensus models such as proof of work (PoW) \cite{Nakamoto2008}, practical Byzantine fault tolerance \cite{Castro2002}, proof of stake (PoS) \cite{Wood2014}, and delegated proof of stake (DPoS) \cite{Schuh2017}. For more details, we recommend the recent review paper \cite{Mingxiao2017}. The principle of PoW is to achieve consensus by computing a mathematical problem. Algorithms that have been used as the mathematical functions include SHA256 \cite{Courtois2014}, Scrypt \cite{Percival2016}, Cryptonight \cite{Van2013}, and Equihash \cite{Biryukov2017}. Miners who want to package new transactions by generating the next block in the blockchain must solve this problem. In addition, a difficulty adjustment algorithm (DAA) \cite{Nakamoto2008} is used to ensure that the time for generating a block remains relatively stable. However, the original Bitcoin PoW model does not have a DAA. In many cryptocurrency settings, the PoW plays an important role in maintaining their consistency.  In this consensus mechanism, blockchain ledger certificates are generated by miners, and each certificate is passed a recent deal. The cryptocurrency network generates a new trading block through which its PoW is validated by the blockchain nodes.

In addition, a blockchain transaction occurs when a user wants to transfer blockchain data (a coin) to another user. In this situation, he/she will send the public key of the targeted user and sign the previous transaction, which is hashed into a fingerprint. Thereafter, this transaction will be verified by a selected miner and simultaneously broadcast to the whole blockchain. Finally, the miner collects valid transactions into a new block. Using the above-mentioned PoW consensus model, a certain user in the blockchain can add a new block to the blockchain, which  will confirm all the transactions packaged in this block until the following six blocks have been verified and added. Usually, the choice of the signature scheme used in the transaction depends on the efficiency, function, and other requirements of the blockchain. For example, the ECDSA \cite{Johnson2001} scheme is used in Bitcoin to achieve more transactions per second (TPS), and a linkable ring signature scheme \cite{Liu2004} is used in Monero to achieve privacy protection. Moreover, the time interval and the size of the blocks are generally fixed. For example, Bitcoin produces a block in 10 min, and each block is 1 MB (Megabyte). The ledger holds each historical transaction (a public key plus a signature) whose size will become larger over time and needs to be expanded. Thus, determining the ways to improve efficiency and expand the blocks are two major obstacles to the transaction part of the blockchain.

However, the above-mentioned cryptographic algorithms underlying the blockchain are not suitable against  adversaries equipped with quantum computers \cite{Aggarwal2017}. The quantum computing attacks on the two major parts of a blockchain include:

\begin{itemize}
	
	\item Attacking the consensus model. For example, a quantum computer will attack a Bitcoin's core hash function named SHA256  to attack its PoW consensus model. According to the research \cite{Grover1996} published in STOC 96, its complexity  will decrease from $\mathcal{O}(N)$ (traditional attack method) to $\mathcal{O}(\pi/4 \sqrt{10N})$.

	\item Attacking the signature schemes. For example, a quantum computer will attack the basic signature scheme, ECDSA, in a Bitcoin's transaction part to double spend the coins. According to the fastest research in \cite{Roetteler2017}, its  complexity is only $9n+2[log_2(n)]$ + 10 with $n$ = 160.
	
\end{itemize}

To this end, the first solution for a post-quantum consensus algorithm came in the form of a new PoW algorithm \cite{Ding2019} proposed  in the conference of applied cryptography and network security in 2019. The author uses a system of quadratic multivariate equations to construct a hash-based hard problem and obtain the solutions (e.g., mining process) by enumeration. However, the mining process of this algorithm does not consider the storage consumption, and thus boils down to a trivial version of Bitcoin's PoW algorithm with a post-quantum feature. On the other hand, the solution to improve the security of transaction processing in the blockchain is called post-quantum blockchain (PQB) \cite{Fernandez2020}, and it currently receives more attention. For example,  Ref.\cite{Yin2018} and Ref. \cite{Li2018} presented a blockchain that enables post-quantum signature schemes for transactions in blockchain systems by using lattice-based cryptography. However, there are other methods that can mitigate quantum attacks, that is, in the consensus part. It is also important to pay more attention to more efficient post-quantum signature solutions. Blockchain has the ability to meet these demands as it enables smart applications to perform data processing.

\subsection{Our Contribution}

In this study, we focus on the above discussion and construct a post-quantum blockchain (PQB), which is a new emerging architecture that can be applied for smart city applications \cite{xie2019survey}, to keep the cities safer and provide a better place of livelihood. Specifically, we present a new post-quantum PoW consensus protocol as the first step. Following the basic transaction construction, we construct a lightweight mechanism using a public key reduced identity-based signature scheme. Our work can support the following properties: both the consensus model and signature scheme mechanisms are post-quantum-resistant, the final blockchain enjoys a novel PoW consensus mechanism with a difficulty adjustment strategy, and we obtain the shortest length of the public key and signature for blockchain transactions.

\textbf{(1) A post-quantum PoW consensus algorithm:} We propose a new post-quantum PoW consensus protocol, which can be used to protect the blockchain under existing quantum computing attacks. As pointed out before, the hash function is not only related to the NP-hard problem but also cannot resist quantum attacks. Because Bitcoin's consensus algorithm uses a hash function, SHA256, that is potentially weak (partially reduced by the quantum attack), our motivation is to use some other quantum-resisted hard problems to replace SHA256. More precisely, our PoW consensus algorithm introduces the problem of solving multivariate quadratic equations, which is an NP-hard problem. In addition, we provide three computing types for solving this hard problem. In addition, the proposed PoW protocol has the following benefits: i) The miner's work includes both computational and storage capabilities. ii) We add dynamic DAAs in our consensus construction.

\textbf{(2) A lightweight PQB transaction protocol:} We propose a new mechanism to construct a lightweight PQB transaction, which embeds a post-quantum identity-based signature scheme and the InterPlanetary  File System (IPFS). In particular, we provide a detailed description of the ways to construct the PQB transaction. The proposed protocol can protect the transaction implementation in the blockchain from potential quantum attacks. In addition, it enjoys high efficiency. For example, the current transaction performance of Bitcoins is 7 TPS. However, if our protocol is deployed into the blockchain system, its performance increases up to 24 TPS in a theoretical analysis.

\subsection{Organization}

The remainder of this paper is organized as follows. Related work is presented in Section \ref{Sec:Related}. Then, in Section \ref{Sec:background}, we introduce the necessary background. The framework of our PQB is presented in Section \ref{Sec:frame}. In Sections \ref{Sec:Consensus} and \ref{Sec:Transaction}, we present our proposed PoW consensus algorithm and lightweight post-quantum blockchain transaction, respectively. Performance evaluation and analysis are provided in detail in Section \ref{Sec:EvaluateResult}. Finally, we present a brief conclusion and future work in Section \ref{Sec:Conclusion}.

\section{Related Work} \label{Sec:Related}

In this section, we review related work on both the blockchain consensus and blockchain transaction parts.

\subsection{Blockchain Consensus}

In 2016, Eyal et al. \cite{Eyal2016} proposed a new consensus algorithm called Bitcoin-NG. The protocol uses two different blocks. The key block is used to elect a leader, and the micro block is used to contain transaction data. The key blocks are generated using the Bitcoin PoW consensus algorithm, and then the leader generates the microblocks for a period of time. The ByzCoin \cite{Kogias2016} consensus algorithm proposed in August of the same year borrowed from this design idea to enable the blockchain system to achieve high performance and low latency beyond Paypal throughput. Elastico \cite{Luu2016} proposed by researchers in 2016 strengthened the scalability of the blockchain through sharding technology. The idea is to isolate the mining network into multiple shards, and different shards process different transaction sets in parallel. The collection of all shards is a complete blockchain ledger. PoET \cite{Buntinx2020} is a random consensus algorithm based on a specific trusted execution environment (TEE). PoET is adopted by the super ledger HyperLedger's sawtooth lake sawtooth project. The basic idea is that each blockchain node generates a random number according to a certain probability distribution to determine the waiting time for generating the next block. TEE can help nodes generate proofs of this waiting time and can be easily verified by other nodes. Owing to PoET, the blockchain system does not require a large amount of power to mine, and it also achieves the fairness of one CPU and one vote. Proof of space (PoSp) proposed in 2014 \cite{Ateniese2014} and Proof of useful work (PoUW) proposed in 2017 \cite{Ball2017} also attempted to solve the energy consumption problem of PoW.  PoSp consensus requires miners to produce a certain amount of storage space (not computing power) to mine, while PoUW replaces the useless SM3  hash operation in PoW consensus with other valuable operations, such as computing orthogonal vector problems, 3SUM problems, and shortest path problems.

In 2013, the Bitshares project proposed a new consensus algorithm, namely DPoS \cite{Schuh2017}. The basic idea of DPoS is that each node in the system can grant its share equity (the currency held and the corresponding currency age) as a vote to a representative; authorizing them to verify and package transactions and produce new blocks. DPoS can not only solve the problem that PoW wastes energy and joint mining poses a threat to the decentralization of the system but can also solve the shortcomings of the participants in PoS who do not necessarily want to participate in accounting. In 2016, the Turing Award winner and MIT professor Sivio Micali \cite{Gilad2018} proposed a fast Byzantine fault-tolerant consensus algorithm called AlgoRand. This algorithm selects the accounting node through the password lottery technique and uses its designed Byzantine fault-tolerant to reach a consensus on the new block. AlgoRand is characterized by simplicity, speed, and a small amount of calculation.

Conflux \cite{Chenxing2018} is a fast, scalable, and decentralized blockchain system that can process concurrent blocks without discarding any branches. The Conflux consensus protocol deals with the relationship between blocks by using a directed acyclic graph (DAG), and the consensus is reached based on the total order of blocks; subsequently, the total order of transactions is deterministically derived from the blocks. For a typical Bitcoin transaction, the throughput is equivalent up to 6400 TPS; thus, the consensus protocol is no longer a throughput bottleneck. Recently, Kim et al. \cite{Kim2019} developed a new shard-based consensus scheme based on a two-phase cooperative game paradigm to provide scale-out system performance. Sharding technology allows the blockchain nodes to obtain only a part of the complete blockchain to improve throughput. According to the idea of egalitarian bargaining solutions, the scheme assigns a transaction set to each shard. Finally, each node can adaptively share rewards by using the concept of proportional bargaining solution.

\subsection{Blockchain Transaction}
Owing to the emergence of Bitcoin, security, and privacy, especially for smart cities \cite{xie2019survey}, many cross-blockchain solutions have been developed. A large number of researchers have proposed dealing with the transaction of various blockchain projects based on its  model. Note that some of them focus on the cross-chain transaction of different blockchains, some focus on privacy protection, and some focus on the expansion of blockchain transactions.

First, we deal with cross-chain transactions. In this field, the notary mechanism is adopted to listen to events in one chain by electing one or more organizations (as the notary) and perform corresponding actions in another chain after an event occurs. In the white paper published by Ripple \cite{Armknecht2015} in 2015, a protocol for the interaction of different blockchain systems is proposed, which suggests that it can be used for the collaboration of blockchain ledger and traditional payment systems. In this protocol, the inter-ledger delivers cross-chain transactions with a connector. Then, various systems run the protocol on the transaction pathway and locks the sender's funds at each system until the final receiver completes the transaction. This trusteeship and execution of cross-chain transactions can be divided into two modes. In one mode, the participants guarantee the security of payment by providing the payment. In the other mode, the participants execute instructions to obtain a payment. In Corda \cite{Brown2016}, the transaction parties jointly select the notary as the cross-verifier and use cryptographic signature schemes to verify the transaction data, which is considered to be a safer notary mechanism. At present, it is considered that cross-chain data transaction verification of the notary mechanism needs to be guaranteed by the signature scheme, which is characterized by low communication efficiency or high consumption of resources. For this reason, scholars advocate the adoption of two extended signature schemes for efficient cross-chain security communication, namely on/off-line signature schemes \cite{Shamir2001} and proxy signature schemes \cite{Mambo1996}.

Second, the objective of privacy protection for blockchain transactions is to prevent malicious nodes from obtaining accurate transaction data. Currently, researchers have proposed various privacy protection schemes for blockchain transactions. One intuitive method is to obfuscate the transaction contents without changing the transaction results. This method is widely used in cryptocurrency; it is called the ``mixed currency'' mechanism. The mechanism of mixed currency originated from an article published by Chaum \cite{Chaum1981}. The primary idea is to use an intermediary to transfer information; thus, an attacker cannot accurately determine the sender or receiver. It can be further improved by using multiple intermediaries to increase the analysis difficulty. Following this paper, there are many studies offering mixed currency services. However, these schemes all require a trusted third party (TTP) that decreases the blockchain's decentralized property. Thereafter, the first decentralized mixed currency method named the CoinJoin mechanism was proposed by Maxwell \cite{Maxwell2017}. The core idea in his paper was to hide the corresponding relationship between the input and output of a transaction by combining multiple transactions into one transaction. It uses transactions with multiple inputs and multiple outputs; thus, a potential attacker cannot effectively distinguish between inputs and outputs by reading the transaction information. Since then, many improvements have been made; for example, Ruffing et al. \cite{Ruffing2014} proposed a completely decentralized Bitcoin mixing protocol named CoinShuffle. They designed an output address shuffling mechanism so that the mixed currency process can be implemented without the need of any TTP, and additionally ensured that the participants did not know the corresponding relationship of other trading parties. However, the CoinShuffle scheme requires participants to be online at the same time, which makes them vulnerable to denial-of-service attacks. To this end, Bissias et al. \cite{Bissias2014} designed a method named Xim to anonymously find participants in mixed coins by utilizing advertising information. Xim is a multi-round hybrid protocol with a manageable success rate. Compared with the CoinJoin mechanism, the cost of attacks in the Xim scheme will increase once the number of participants in mixed currency increases; thus, it is effective to resist denial-of-service attacks. Thereafter, Ziegeldorf et al. \cite{Ziegeldorf2015} adopted a secure multi-party computing protocol to propose an improved method named CoinParty, which can be used in the case of malicious operations from some mixed nodes. Monero designed by Shen et al. \cite{Shen2015} adopts a ring signature mechanism to improve the process of mixing coins, while Zcash \cite{Ben2013} uses zero-knowledge proof technology to verify the transaction. The proof of process  of these two methods need not reveal relevant information; therefore, they can hide the sender and receiver of blockchain transactions and even the data of the transaction.

Finally, the expansion of data transactions is mainly referred to by modifying the data block or increasing the blocks to accommodate the number of transactions. The related techniques are as follows: expanding block, segregated witness, and DAG. For example, Bitcoin Cash \cite{BCH2018} can be increased from the original 1 MB block to 8 and 32 MB blocks successively, by expanding its block. Increasing the block size is the simplest, most straightforward way, but the size of the block cannot be arbitrarily expanded. An increase in the block size requires a corresponding increase in the processing capacity, which will lead to a problem of centralization of computing power, and thus is more vulnerable to external attacks \cite{Decker2013}. To deal with the expansion problem of blockchain systems, another common technique, segregated witness (SegWit) \cite{Wuille2015}, is proposed. In this technique, a blockchain can use a segregation witness to reduce the transaction contents of the verification data. However, the SegWit mechanism merely changes the data structure of the transaction data, and the verification data essentially still occupies the storage of the block, and thus the blockchain cannot be released. Another expansion method is to use DAG to change the linear storage structure of the blockchain \cite{SERGIO2019}. In a DAG coin, each new transaction can be submitted to a separate ``block.''  The consensus mechanism of DAG is no longer the traditional broadcast data verification method, but a hash value of the previous block is transferred according to certain rules. The record  storage of data can be parallelized, thus improving the throughput of the block chain network.

\section{Background} \label{Sec:background}

\subsection{Proof of Work Model in Blockchain}

In general, a consensus algorithm is established in the blockchain ecosystem to decide whether to add an entry (or block). It uses the following principles to generate blocks and maintain the blockchain:

\begin{itemize}
	
	\item Adopt the consensus algorithm to ensure there are no malicious nodes.
	
	\item All nodes implement the consensus mechanism, generate blocks, and reward the winners.
	
	\item The longest chain formed by all the generated blocks is a legal blockchain.
	
\end{itemize}

\begin{figure}[h]
	\center
	\includegraphics[width=3.8in]{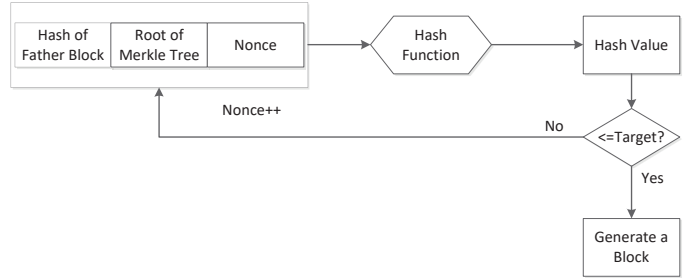}
	\caption{Workflow of Bitcoin PoW model \label{fig:fig141}}
\end{figure}

\begin{figure*}[ht]
	\center
	\includegraphics[width=4.3in]{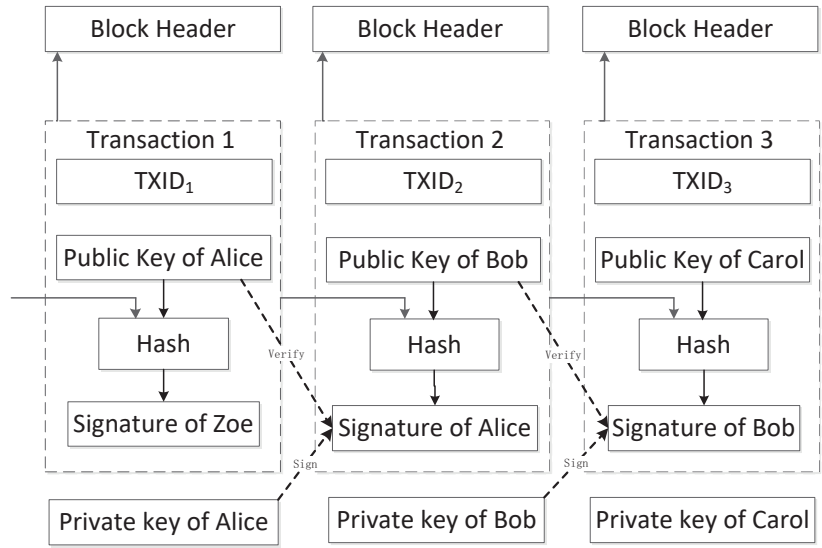}
	\caption{Transaction in Bitcoin \label{fig:fig1}}	
\end{figure*}

In this paper, we mainly focus on the famous PoW consensus model, especially Bitcoin's PoW model. We summarize the workflow of Bitcoin's PoW model in Fig. \ref{fig:fig141}.

In the PoW model, users publish the next block by first solving a computationally substantial mathematical problem. The answer to this problem is the ``proof" of their work. This problem is designed in such a way that solving the problem is difficult, but verifying that a solution is satisfactory is easy \cite{Ding2019}. This allows all other blockchain nodes to easily verify the next block of any proposal, and any proposed block that does not satisfy a puzzle will be rejected.

A common puzzle for the PoW model is that it requires a hash digest of the block header to be less than a certain difficulty target value. Participant nodes make many small changes to their block headers (for example, changing the value of nonce) in an attempt to find a hash value that meets the requirements. For each attempt, the publishing node must calculate the hash of the entire block header. Hashing the block header multiple times becomes computationally intensive. The target value may be modified over time to adjust the difficulty (up or down) to satisfy the required frequency of block releases.

\emph{Remark.} Bitcoin's PoW uses only the computation capability (no memory capability) of a mine, and is not flexible (which makes the BITMAIN company successful after the promotion of their Antminer $-$ASIC chip products). In addition, there is no DAA in the original Bitcoin PoW model. It simply adjusts the difficulty for every 2016 blocks and averages the block creation rate such that it is approximately once per ten minutes.

\subsection{Transaction in Blockchain} \label{Transaction}

This subsection presents the definition of the access structure. The core security problem in the blockchain transaction is preventing double spending attacks \cite{Budish2018}.

The transaction process is shown in Fig. \ref{fig:fig1}. There are two flows named transaction inputs and transaction outputs in every transaction, and all transactions constitute a chain structure. Assume a user, Alice, wants to transfer 1 Bitcoin to Bob; Alice will first apply for a transaction, and then send a request information (Transaction 2 in Fig. \ref{fig:fig1}) to all the blocks in the whole blockchain network. All blocks are then given account balance and transfer information for both Alice and Bob and verified. In particular, as shown in Fig. \ref{fig:fig1}, Alice transfers the coin to Bob by signing the hash of Transaction 1 and Bob's public key using her private key. Other miners can verify the signature of Transaction 2. Bob has ownership of the transferred Bitcoin. When Bob wants to spend this coin, he can use his private key to generate Transaction 3 in the same way as above.

\subsection{Algebraic Algorithms for Solving the System of Quadratic Equations} \label{Grobner }

The basic puzzle objects of our new PoW algorithm are a system of multivariate quadratic equations with $m$ equations in $n$ variables over a finite field \({F}_q\). Such a system of $m$ questions in $n$ variables is defined as
\begin{equation}
\left\{ \begin{array}{l}
\sum\limits_{i = 1}^n {\sum\limits_{j = i}^n {\alpha _{ij}^{(1)} \cdot {x_i} \cdot {x_j}} }  + \sum\limits_{j = i}^n {\beta _i^{(1)} \cdot {x_i} + } \gamma _i^{(1)} = 0;\\
...\\
\sum\limits_{i = 1}^n {\sum\limits_{j = i}^n {\alpha _{ij}^{(m)} \cdot {x_i} \cdot {x_j}} }  + \sum\limits_{j = i}^n {\beta _i^{(m)} \cdot {x_i} + } \gamma _i^{(m)} = 0.
\end{array} \right.
\end{equation}

Such systems can be solved by algebraic algorithm techniques such as XL \cite{Courtois2000} and Gr\(\rm{\ddot o}\)bner basis algorithms such as F$_4$ \cite{Faugere1999} and HF$_5$ \cite{bettale2009hybrid}. To summarize, the best direct attack algorithm is the algorithm HF$_5$ for large fields, and the algorithms F$_4$ for medium fields to solve multivariate polynomial equations.  When the finite field is small or medium and the number of equations is much larger than that of the variables, the best algorithm is the XL algorithm.

\begin{table*}[htb]
	\caption{Algebraic algorithms for solving the system}
	\label{table_1}
	
	\centering
	\begin{tabular}{c| c| c}\hline
		\centering
		\textbf{Parameter situation} & \textbf{Algorithms} & \textbf{Theoretical}  \textbf{complexity}  \\\hline
		Medium fields (i.e. $q=2^{16}$) &F$_4$ &\(\mathcal{O} \left( m{{{\left( {\left( \begin{array}{l}
						(m+n) + {d_{reg}} - 1 \\
						{d_{reg}} \\
						\end{array} \right)} \right)}^\omega }} \right)\)   \\\hline
		Large fields (i.e. $q=2^8$) &HF$_5$ & \(q^k \mathcal{O}  \left(m {{{\left( {\left( \begin{array}{l}
						(m+n)-k +  {d_{reg}} - 1 \\
						{d_{reg}} \\
						\end{array} \right)} \right)}^\omega }} \right)\)   \\\hline
		Small/medium field and overdefined system  & & \\
		(i.e. $ m= \varepsilon n^2, \varepsilon > 0$) &XL  &\(\mathcal{O}({(n)^{\omega D}}/D!)\)  \\\hline
	\end{tabular}
	\flushleft
	
	Notation for Table \ref{table_1}: $d_{reg}$ is the degree of regularity of the multivariate system,  $\omega$ is a linear algebra constant, and $2 \le \omega \le 3$; $k$: a fixed constant for F$_5$; $q$: the underground finite fields;  $D$: \(D \approx \left\lceil {1/\sqrt \varepsilon  } \right\rceil \).
	
\end{table*}

According to references \cite{bardet2004complexity}, \cite{bettale2009hybrid}, and \cite{Courtois2000}, the complexity of all three algorithms is summarized in Table \ref{table_1} . Thus, in the construction of our PoW algorithm, we use the algorithm F$_4$ as our recommended algorithm.

\subsection{ID-Rainbow}

An ID-based signature scheme consists of the following algorithms: \textsf{Setup}, \textsf{Extract}, \textsf{Sign}, and \textsf{Verify}.

\begin{itemize}
	\item \textsf{Setup}: On an input of a security parameter $k$, it produces the master secret key $msk$ and the common master public $mpk$, which includes a description of a finite signature space and a description of a finite message space.
	
	\item  \textsf{Extract}: On input of the signer's identity $ID \in \{0, 1\}^*$ and the master secret key $msk$, it outputs the signer's secret signing key $usk_{ID}$. 
	
	\item  \textsf{Sign}: On input of a message $m$, a user's identity $ID$, and the secret keys of one member $usk_{ID}$, it outputs an ID-based signature $\sigma$ on the message $m$.
	
	\item  \textsf{Verify}: On input of a signature $\sigma$, a message $m$, and the signers' identity $ID$, it outputs $1$ for \textit{true} or $0$ for \textit{false}, depending on whether $\sigma$ is a valid signature signed by a certain member on a message $m$.
	
\end{itemize}

Recently, Chen et al. \cite{Chen2019} proposed a post-quantum identity-based signature scheme named ID-Rainbow based on Rainbow \cite{Ding2005}, which makes the public key relate to the user's identity information, and results in a public key length of 8 bytes with a security level of 80 bits. For more details, we recommend reading Ref. \cite{Chen2019}.

The main idea of their work is to embed the user's id into every coefficient of the secret keys and public keys and then construct a common Rainbow signature. Below, we summarize the process to embed the user's id into the secret key and public key of Rainbow.

Let the user's id is denoted by $ID(U_{i})$ = $(z_{1}$, $z_{2}$, $\dots, z_{d})$ and $L_{1}(x_{1}$, $\dots$, $x_{m})$ = $(L_{1,1}(x_{1}$, $\dots$, $x_{m})$, $\dots$, $L_{1,n}(x_{1}$, $\dots$, $x_{m}))$, where $L_{1,i}(x_{1}$, $\dots$, $x_{m})$ = $\sum L_{(1,i,j)}(z_{1}$, $\dots$, $z_{d})x_{j}$ + $L_{(1,i,0)}$$(z_{1},$ $\dots, z_{d})$. Note that each $L_{(1,i,j)}(z_{1}, \dots,z_{d})$ is a linear function of $z_{1}, \dots, z_{d}$. For example, let $m$ = 2, $d$ = 2. We have $L_{1}(x_{1},x_{2})$ = $((z_{1}+2z_{2})x_{1}$ + $(2z_{1}-z_{2})x_{2}$, $(z_{1}-z_{2})x_{1}$ + $z_{1}x_{2})$. Also let $L_{2}(x_{1}$, $\dots$, $x_{n})$ = $(L_{2,1}(x_{1}$, $\dots$, $x_{n})$, $\dots$, $L_{2,n}(x_{1}$, $ \dots$, $x_{n}))$, where $L_{2,i}(x_{1}$, $\dots$, $x_{n})$ = $\sum L_{(2,i,j)}(z_{1}$, $\dots$, $z_{d})x_{j}$ + $L_{(2,i,0)}(z_{1}$, $\dots$, $z_{d})$. Thus,
$F_{l}$ = $\sum_{i,j}\alpha_{l_{ij}}x_{i}x_{j}$ + $\sum_{i}\beta_{l_{i}}x_{i}$ + $\gamma_{l}$, where $\alpha_{l_{ij}}$ = $A_{l_{ij}}(z_{1}$, $z_{2}$, $\dots $, $z_{d})$, $\beta_{l_{i}}$ = $B_{l_{i}}$$(z_{1}$, $z_{2}$, $\dots$, $z_{d})$, and $\gamma_{l}$ = $C_{l}(z_{1}$, $z_{2}$, $\dots$, $z_{d})$, are all linear functions of \{$z_{1}$, $\dots$, $z_{d}$\}.

Based on the above idea, the four algorithms of ID-Rainbow are as follows:

\begin{itemize}
	\item  \textsf{Setup}: Given the system parameter, the scheme takes polynomials with coefficients that are used to compute the master public key, that is, $mpk$ = $({\overline \alpha  _{lij}}$, ${\overline \beta _{li}}$, ${\overline \gamma  _l})$. The according master secret keys are polynomials with coefficients $(L_{(1,i,j)}$, $L_{(1,i,0)}$, $L_{(2,i,j)}$, $L_{(2,i,0)}$, ${\alpha _{lij}}$, ${\beta _{li}}$, ${\gamma _l}$, and the specific coefficients $L_1'$ and $L_2'$ corresponding to every user. More precisely, the master secret key is $msk $ = $(L_{(1,i,j)}$, $L_{(1,i,0)}$, $L_{(2,i,j)}$, $L_{(2,i,0)}$, ${\alpha _{lij}}$, ${\beta _{li}}$, ${\gamma _l}$, $L_1'$, $L_2')$. Finally, the master public key will be publicly known, while the master secret key will be known only to the key distributed center (KDC).
	
	\item  \textsf{Extract}: Given an arbitrary identification of a specific user $ID_u$ = \{$z_{1}$, ..., $z_{d}$\}, KDC can compute the public polynomials $P$ by using the master public key $mpk$. In addition, by using the master secret key $msk$, KDC can compute the private key $F$, $L_1'$, $L_2$, and $L_1'$, $L_2'$ via this $ID_u$. Then, the KDC will extract the private key of a specific user as $L_{1u} $ = $L_1 \circ $ $L_1'^{-1}$, $F_u $ = $L_{1'} \circ F \circ L_2'$, $L_{2u}$ = $L_2'^{-1} \circ L_2$. Finally, this private key is distributed to this specific user $ID_u$.
	
	\item  \textsf{Sign}: Given the private key $L_{1u}$, $F_u$, $L_{2u}$, and the message $M$, the scheme returns $X$ = \{ ${x_1}$, ..., ${x_{o + v}}$\} as the legitimate signature by running a regular Rainbow signature process.
	
	\item  \textsf{Verify}: To verify the signature $X$ of document $M$ on input the public key $ID$ = $z_{1}$, ..., $z_{d}$, the verifier substitutes $X$ to the public master public key $mpk$ and computes the public polynomials $P$. Then, in the verification process of ID-Rainbow, the scheme checks if $P(X) $ = $M$, this signature is valid once it is true. Otherwise, it's not.
	
\end{itemize}

\section{Framework of Post-quantum Blockchain} \label{Sec:frame}

\begin{figure}[h]
	\center
	\includegraphics[width=3.1in]{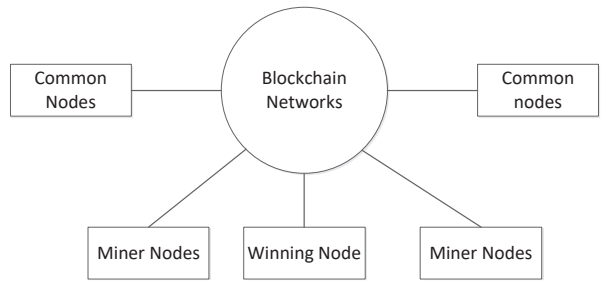}
	\caption{Framework \label{fig:Framework}}
\end{figure}

The framework of our designed blockchain system, as shown in Fig. \ref{fig:Framework}, includes: blockchain network, wining node, miner nodes, and common nodes. The details are presented below.

\textit{\textbf{Blockchain network}}: it is used for constructing a real blockchain system, and winning node, miner nodes, and common nodes are considered as blockchain nodes. In a blockchain system, the blockchain nodes can process the transaction service in response to a transaction request; thereafter, the transaction will be broadcast to the blockchain network after the transaction is successfully created. Each transaction includes transaction data, that is, the 10 BTC from user $A$ to user $B$, and verification data, which is used to verify whether the transaction data is legal, that is, verification data includes the public key and the signature of user $A$.

\textit{\textbf{Miner nodes}}: in our framework, the miner nodes can receive a number of transactions over a period of time; each transaction includes transaction data and transaction verification data. The role of the miner nodes is to confirm a transaction and package multiple transactions into blocks. After the transaction is broadcast to the blockchain network, the miner nodes respond to generate multiple transactions over a period of time.

\textit{\textbf{Winning nodes}}: it is a special miner none which is regarded as the node who generate a novel block in the blockchain system.

\textit{\textbf{Common nodes}}: finally, the common nodes are used for dealing with the extra data which will be discussed in the following sections.

\section{Our New Post-quantum PoW Consensus Model} \label{Sec:Consensus}

\begin{figure}[htb]
	\center
	\includegraphics[width=3.8in]{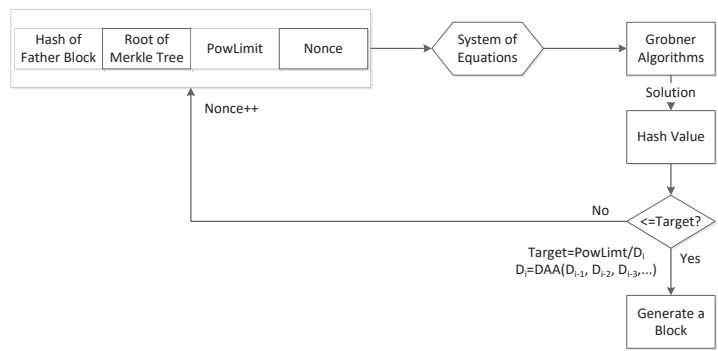}
	\caption{Workflow of our PoW model \label{fig:fig11}}
	
\end{figure}

In this section, we propose a new PoW model. This model is run by the entity of the winning node, miner nodes, and the blockchain network, and the workflow of our PoW model is shown in Fig. \ref{fig:fig11}.

The main idea behind the construction of such a PoW model is to use some other hard problems to replace SHA256 using Bitcoin's PoW model.  More precisely, our PoW consensus algorithm introduces the problem of solving multivariate quadratic equations, which is an NP-hard problem.  We also provide three parts' behavior for the blockchain network, miner nodes, and the winning node, as described in the following sections.

\subsection{The Blockchain Network}

The Blockchain network will perform the following processes as shown in Algorithm \ref{OurGen}:

\begin{algorithm}[htb]
	
	\caption{ \textsf{Blockchain Network's Running Algorithm}}
	\label{OurGen}
	
	\begin{algorithmic}[1]
		\STATE Construct a block structure of the blockchain network;
		\STATE Set the calculated minimum POWLimit of the consensus algorithm;
		\STATE Set the dynamic DAA;
		\STATE Set the selected Gr\(\rm{\ddot o}\)bner solving algorithm, for example, to use the F$_4$ \cite{Faugere1999} algorithm;
		\STATE Set the target value in the same way as the PoW algorithm, which will be used to verify the mining effectiveness of the mining nodes;
		\STATE Generate a block for the winning node and give it a reward.
	\end{algorithmic}
\end{algorithm}

\begin{figure}[htb]
	\center
	\includegraphics[width=3.9in]{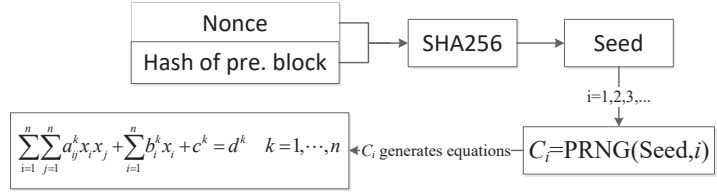}
	\caption{Generating a system of equations \label{fig:equations}}
\end{figure}

Furthermore, the process of generating a system of multivariate polynomial equations is adopted in the workflow, as shown in Fig. \ref{fig:equations}. Here, the number of random values, $C_i$, is $n^2/2 \times m+n \times m+2m$, and it is used to construct a system of multivariate polynomial equations.

Specifically, the father block and the random number are hashed as seeds, then the coefficients of the multivariate polynomial equations are generated by the pseudo-random number generator according to the seeds, and finally a system of equations composed of $n$ variables and $m$ equations is generated. The number of random values is determined according to the scale of the equation system (herein, we calculate as $n^2$ $\times$ $m$ + $n$ $\times$ $m$ + $2m$). More precisely, to fix this set of random equations, we can observe that the number of random coefficients is the sum of the number of quadratic term coefficients, number of first order coefficients, and number of constant coefficients on the right side of the equation; therefore, the number of random numbers required is $n^2$ $\times$ $m$ + $n$ $\times$ $m$ + $2m$.

Finally, the generated system of equations $F$ = $(f_1$, ..., $f_m)$ is
\begin{equation}
\left\{ \begin{array}{l}
\sum\limits_{i = 1}^n {\sum\limits_{j = i}^n {\alpha _{ij}^{(1)} \cdot {x_i} \cdot {x_j}} }  + \sum\limits_{j = i}^n {\beta _i^{(1)} \cdot {x_i} + } \gamma _i^{(1)} = 0;\\
...\\
\sum\limits_{i = 1}^n {\sum\limits_{j = i}^n {\alpha _{ij}^{(n)} \cdot {x_i} \cdot {x_j}} }  + \sum\limits_{j = i}^n {\beta _i^{(n)} \cdot {x_i} + } \gamma _i^{(n)} = 0,
\end{array} \right.
\end{equation}
where all the coefficients are related to the random value $C_i$ with $i$ = $1, \dots, n^2 \times m+n \times m+2m$.

Now, we face solving the system of $m$ questions in $n$ variables, which are defined as
\begin{equation}
\left\{ \begin{array}{l}
\sum\limits_{i = 1}^n {\sum\limits_{j = i}^n {\alpha _{ij}^{(1)} \cdot {x_i} \cdot {x_j}} }  + \sum\limits_{j = i}^n {\beta _i^{(1)} \cdot {x_i} + } \gamma _i^{(1)} = 0;\\
...\\
\sum\limits_{i = 1}^n {\sum\limits_{j = i}^n {\alpha _{ij}^{(m)} \cdot {x_i} \cdot {x_j}} }  + \sum\limits_{j = i}^n {\beta _i^{(m)} \cdot {x_i} + } \gamma _i^{(m)} = 0.
\end{array} \right.
\end{equation}

Note that we can use the Gr\(\rm{\ddot o}\)bner basis algorithm F$_4$  \cite{Faugere1999} or HF$_5$ \cite{bettale2009hybrid} to solve it, while the parameters $(m,n)$ are appropriate.

\subsection{The Miner Nodes and the Winning Node}

Miners use the Gr\(\rm{\ddot o}\)bner basis solution algorithm (F$_4$, F$_5 $) to solve the random quadratic multivariate equations generated based on the seed number as described above, and output the solution \textit{Solution} = $(x_1$, $x_2$, ..., $x_n)$. More precisely, after generating a finite quadratic multivariate equation system $F$ = $(f_1$, ..., $f_m)$ , all the miner nodes will use the Gr\(\rm{\ddot o}\)bner basis solution algorithm F$_4$ or F$_5$ to solve $(x_1$, $x_2$, ..., $x_n)$ = \textit{Gr\(\rm{\ddot o}\)bnerBasis}$(f_1,...,f_m)$.

In particular, the miner nodes (including the winning node) in the blockchain system will perform Algorithm \ref{MinerALG}.

\begin{algorithm}[htb]
	
	\caption{\textsf{Mining Algorithm of the Miner Nodes}}
	\label{MinerALG}
	
	\begin{algorithmic}[1]
		\STATE Check that if the hash value \\
		\textit{SHA256}$(x_1,x_2,...,x_n) \le PowLimit/D_i$ then
		\STATE  Broadcast a new block named \textit{Block}$_i$;
		\STATE Else do the following calculation:\\
		(\textit{Nonce} = \textit{Nonce}+1)\\
		(\textit{Seed} = \textit{SHA256}(\textit{SHA256}(\textit{Block}$_{i-1})$ | \textit{Nonce}))\\
		(\textit{Random}$_i$ = \textit{PRNG}($Seed, i)/q)$;
		\STATE   Use  \textit{Random}$_i$ to construct a new system of equation $F$ = $(f_1$, ..., $f_m)$;
		\STATE   Get a solution using fixed Gr\(\rm{\ddot o}\)bner algorithm (i.e., F$_4$) such that $(x_1,x_2,...,x_n)$ = F$_4(f_1$, ..., $f_m)$.
	\end{algorithmic}
\end{algorithm}

The current miner node outputs \textit{Solution} = $(x_1$, $x_2$, ..., $x_n)$ according to the solution of the quadratic multivariate equations in the finite field, and verifies whether the PoW condition is met. If it is met, a new block will be released, and the current miner node becomes a winning node. It is broadcast to other nodes in the blockchain network. Otherwise, the current block \textit{Nonce} is incremented to generate a new seed value and generate a new random quadratic multivariate equation system. The above process is repeated until any node finds a solution that satisfies the condition \textit{SHA256}($x_1,x_2,...,x_n$) $\le PowLimit/D_i$.

Then, the other miner nodes in the network receive the broadcast by the winning node, they verify whether the \textit{Nonce} and \textit{Solution} in the block header meet the condition, and at the same time verify the legality of each transaction in the block. If both are satisfied, the block is linked to the local blockchain as the latest block, and this block is also broadcast to its neighbors.

The verification algorithm is shown in Algorithm \ref{VerifyALG}.

\begin{algorithm}[htb]
	
	\caption{ \textsf{Verification Algorithm of the Miner Nodes}}
	\label{VerifyALG}
	
	\begin{algorithmic}[1]
		\STATE Generate a seed based on the hash value of the previous block and $Nonce$ as follows: \\
		(\textit{Seed} = \textit{SHA256}(\textit{SHA256}($Block_{i-1}$) | \textit{Nonce});)
		\STATE Generate a system of equations based on the seed $Seed$ by calculating:\\
		(\textit{Seed} = \textit{SHA256}(\textit{SHA256}(\textit{Block}$_{i-1}$) | \textit{Nonce}))\\
		(\textit{Random}$_i$ = \textit{PRNG}($Seed, i)/q$;)
		\STATE Use \textit{Random}$_i$ to construct a new system of equations $F$ = $(f_1, \dots, f_m)$;
		\STATE Substitute the value of \textit{Solution} into the system to verify whether the solution meet all the equations;
		\STATE Generate a block for the winning node and give it a reward if the above is correct and abort;
		\STATE Reject this block and abort.
	\end{algorithmic}
\end{algorithm}

\section{A Lightweight Post-quantum Blockchain Transaction }  \label{Sec:Transaction}

In a blockchain system, the ledger is stored among the nodes, and the historical transaction is stored in a certain ledger. As time passes, the volume of the ledger will gradually expand. On the other hand, the time interval of the block generation and the size of the blocks are fixed, that is, one block generation period of Bitcoin is 10 min, and each block is at most 1 MB, which limits the transaction capacity of each block. For example, the transaction rate of Bitcoin's entire network is only 7 per second, which is not enough for a high transaction concurrency scenario such as the double 11 promotion day of Alibaba. Therefore, a lightweight blockchain transaction mechanism is a very important research direction for the current blockchain.

\subsection{Our Post-quantum Transaction Mechanism}

Equipped with the identity-based Rainbow signature scheme mentioned above, we now construct a post-quantum blockchain-enabled transaction mechanism. We highlight here that our transaction mechanism is lightweight, more precisely, as far as we know, the lightest one. This mechanism runs using the entities such as the miner nodes, common nodes, and blockchain network. And it is shown in Fig. \ref{fig:fig12}. Then, following the description of blockchain transaction in Section \ref{Transaction}, the transaction can be implemented in the following three processes:

\begin{figure*}[htb]
	\centering
	\includegraphics[width=4.3in]{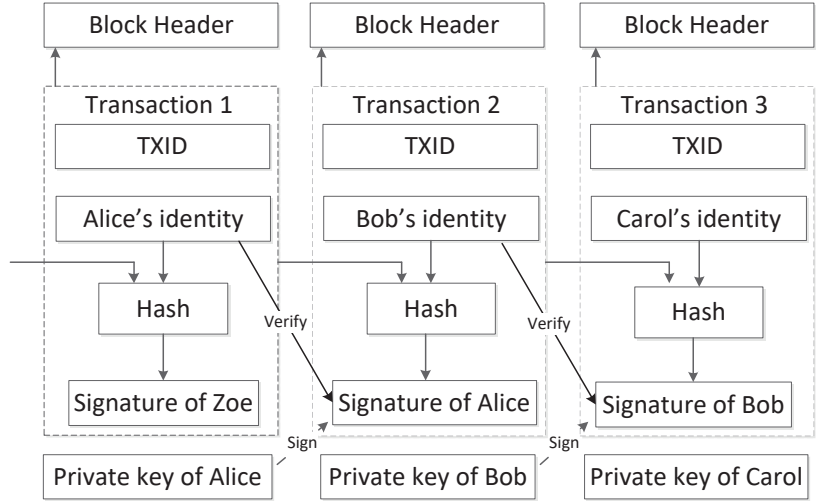}
	\caption{Post-quantum transaction solution in Blockchain system \label{fig:fig12}}
	
\end{figure*}

\begin{figure}[htb]
	\centering	
	\includegraphics[width=4.7in]{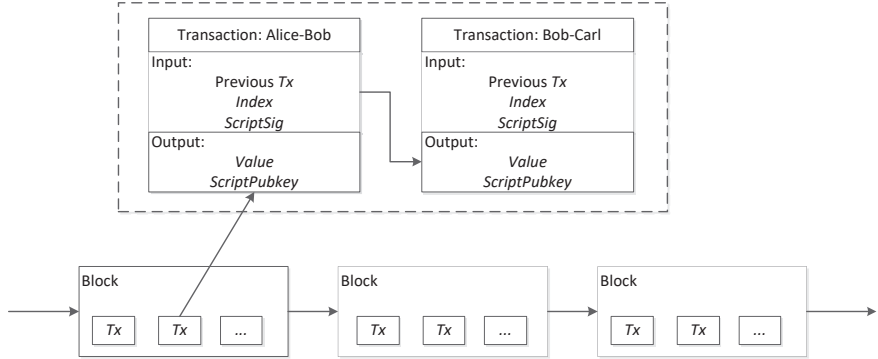}
	\caption{Data structure in blockchain system \label{fig:Data}}
\end{figure}

\textbf{1. Preparation process}. The common nodes or miner nodes are independent entities in the blockchain networks, and they combine and perform different services, that is, transactions. The transaction in fact is a data structure, as shown in Fig. \ref{fig:Data}, and is indexed by a transaction identity (TXID). Generally, the input of a transaction contains the previous \textit{Tx}, \textit{Index}, and \textit{ScriptSig}, where \textit{Tx} stands for a hash value of the previous transaction, \textit{Index} is the value index of the output of the previous node, and \textit{ScriptSig} is the signature of the transaction owner. Finally, the output consists of a \textit{Value} and \textit{ScriptPubkey}, where \textit{Value} is the value of the transaction, and \textit{ScriptPubkey} is the receiver's public key.

Here, the transaction \textit{Index} is used to generate an identification of the user (Alice's identity in Fig. \ref{fig:fig12}), and it can also be used to compute the public key of ID-Rainbow. To resist statistical attacks, a new TXID will generate a new transaction from a different public key. Therefore, we should store more public and private key pairs for new transactions of each user in the blockchain network. To this end, we can use a lightweight wallet designed in Ref. \cite{Chen2019}, which only needs to store the root key. Reducing wallet redundancy is more suitable for transactions implemented on the blockchain.

\begin{figure}[htb]
	\centering
	
	\includegraphics[width=3.6in]{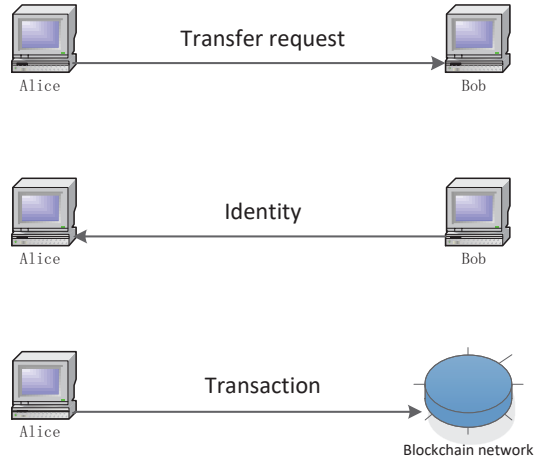}
	\caption{Transaction implementation in blockchain system \label{fig:Implementation}}	
\end{figure}

\textbf{2. Implementation process}. Here, we consider a real scenario; as the common nodes, a user Alice wants to send some data (bitcoins) through a transaction to a user Bob. Then, three steps are executed to accomplish this transaction (as shown in Fig. \ref{fig:Implementation}). First, the Alice sends a transfer request to Bob. Second, Bob generates an address according to his identity number. It  also uses the ID-Rainbow scheme to extract a private key and sends it to Alice for further transaction implementation. Alice then creates and broadcasts transactions across the network for further transactions. It must be emphasized that the total input to the transaction must be equal to the total amount that Alice has. If the required number of transactions Alice sends is more than the total amount, Alice must create a new TXID address to send insufficient bitcoins. Finally, in the transaction execution, the information of the minor node and the reward for creating a new block are recorded in a temporary block. This content is broadcast throughout the blockchain network in the last time period. Once a mining node gains the right to create a new block, the temporary block becomes a common transaction as a compensation.

\textbf{3. Confirmation process}. Note that the transaction confirmation of our PQB is similar to that of Bitcoin. When the transactions are broadcast to the network and verified by the miner node, they will be collected and packaged into a temporary block. Then, when the latest block in the current period is created, the temporary block will be packaged into this new block. In addition, by appending the new block to the longest chain, all transactions in this block are verified once. Subsequently,  the transactions in this temporary block along with the new following block will be verified multiple times (depending on how many following blocks are created during this time period) because this temporary block is now the previous block. We fix the number of following blocks to six. Generally, after six blocks, these transactions cannot be modified because rebuilding six blocks requires many calculations. At this point, the transaction has been stored in the blockchain as an unchangeable record.

\subsection{The Lightweight Post-quantum Segregation Witness Transaction Mechanism}

To deal with the expansion problem of blockchain systems, the current common technique includes segregated witness (SegWit) \cite{Wuille2015} . A blockchain can select a segregation witness that can reduce the transaction validity of the transaction validity verification data. However, the SegWit mechanism merely changes the data structure of the transaction data and the verification data for the validity of a transaction essentially still occupies the storage of the block, and thus cannot release the blockchain in essence. Therefore, in this section, we will improve the current isolation witness scheme to truly realize the expansion of the blockchain account system.

In our segregated witness transaction mechanism, there are two types of data: transaction data and verification data.  The miner nodes first package the transaction data from the transaction users into different blocks, then package the verification data into an extension block associated with the block, and broadcast the blocks and the extended blocks. After receiving the transaction data broadcasted by the user, the miner node verifies the validity and authenticity of the transaction. Here, validity implies that the miner nodes will verify whether the payer's token is sufficient. The miner nodes inquire into the number of tokens transferred to the account in the past legal transaction according to the address of the payer in the transaction. When the amount is greater than or equal to the amount filled in the transaction statement, the transaction is legal. Then, the miner node needs to start a new hash calculation with different random numbers, until a random number matching the target value is found. If a random number is found, the miner node packs the transaction data into blocks, which will be transferred later. In addition, each verification data is moved into the expansion block and is handled outside the block. Finally, only the transaction data is stored in the block.

After generating the block and the extended block, the miner node broadcasts the block and the extended block to the entire network to tell other blockchain nodes that a new block has been generated. When the common node receives the block and the extension block, the transaction data in the block is verified by each relative verification data in the extended block. In our mechanism, a common node is a node that need not provide a query for a complete transaction data service, that is, a node operated by an ordinary individual user, a block browser, or a mining pool. After receiving the block and the extended block, the common node uses the data in the extension block part to verify the validity of the transaction.

\begin{figure*}[htb]
	\centering
	\includegraphics[width=4.3in]{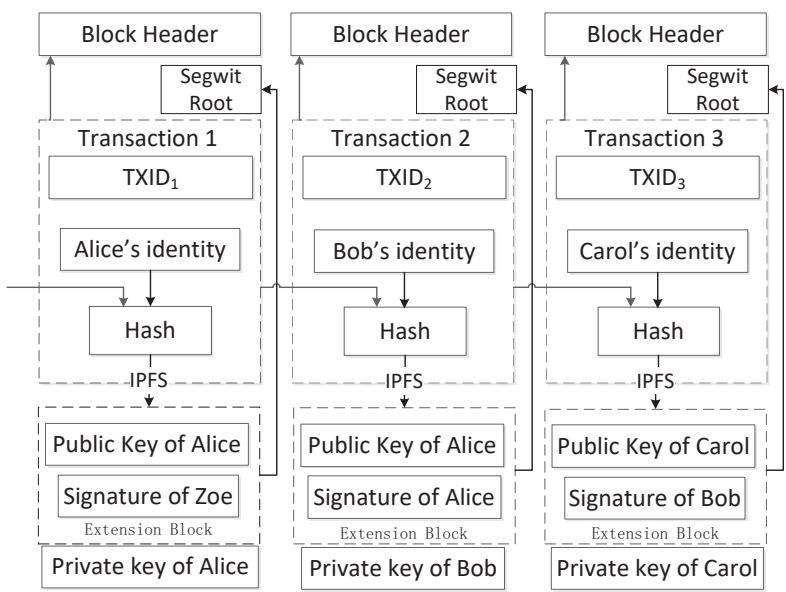}
	\caption{Lightweight transaction in Blockchain system \label{fig:fig122}}
	
\end{figure*}

To verify transaction data, it is necessary to confirm the correspondence between the transaction data in the block and the verification data in the extension block. Therefore, the block records the correspondence between the transaction data in the block and the verification data in the extension block. As shown in Fig. \ref{fig:fig122}, the block records such correspondence by using a hash pointer and embedded with a IPFS storage (i.e., a hash in the figure).

After receiving the block and extension block, the common node verifies the transaction data in the block by using each relative verification data in the extension block. The process includes:

\begin{itemize}
	
	\item After receiving the block and extension block, the common node reads a transaction data in the block.
	
	\item According to the correspondence, the common node confirms that the transaction data corresponding to the relative verification data in the extension block is legal.
	
	\item The common node verifies the transaction data according to the relative verification data.
	
	\item If the verification is valid, the common node reads another transaction data in the block, and verifies transaction data according to the relative verification data until all the transaction data are read.
	
	\item If the verification is valid, the common node loads this block into the local blockchain and discards the extended block. If the verification is invalid, the common node rejects the transaction and discards the data in the extended block.
\end{itemize}

\section{Analysis, Comparison, and Performance Evaluation}  \label{Sec:EvaluateResult}

Because our work is still in its preliminary stage, we do not construct a real cryptocurrency to deploy our PoW consensus algorithm and transaction algorithm. In this section, we present the theoretical analysis and simulation experiments to support the proposed memory mining property of our PoW algorithm and compare the transaction benefits of our lightweight transaction protocols.

\subsection{Security Analysis}

For the security of the PoW schemes that we design, we have to take care of the following attacks.

The first potential attack on the mining process is in fact finding a birthday paradox that can cause collisions on \textit{SHA256}, that is, to find two vectors $V_1$ and $V_2$ with the same hash, and thus pass the verification process. More precisely, once we find a collision, we can reconstruct the core system of equations. Without generation, let $2^z$ be the security of our PoW model. For the hash function \textit{SHA256}, the number of bits of the output of \textit{SHA256} is $E$ = 256.  Let us assume that the hash function is computed in $2^h$ operations.

The birthday attack proceeds as follows: we can compute $2^{z-h}$ values $H(V)$ and store $V$ at the address $H(V)$. Then, with high probability, we will obtain that two values share the same address. In other words, we find a collision $H(V_1)$ = $H(V_2)$ when $z -h$ $> E/2$. Therefore, to avoid this collision attack, we must always have $z>E/2+h$ = 128 + $h$.

The hash function \textit{SHA256} has a speed of approximately 13 cycles per byte. Then, for a block header, such as a Bitcoin of 80 bytes, we will need  $2^{10}$ cycles. Thus, $h$ = 10 for modern computers. This comes from the birthday paradox, with an attack of time complexity $2^{138}$.

Second, another attack is to directly reconstruct the system of equations. As mentioned before, the number of random values, $C_i$, is $n^2 \times m+n \times m+2m$. The structure security is protected by using PRNG; thus, we assume that only enumeration attacks work. To reconstruct such system of equations, $q^{n^2 \times m+n \times m+2m}$.

Thus, even for small potential parameters of $n$ = 12, $m$ = 12, and $q$ = 2, the time complexity of this attack is $2^{2032}$. Finally, we can conclude that the security of our PoW scheme is over $2^{138}$.

\subsection{Simulation Results and Parameter Recommendation for the PoW Model}

In this section, we concentrate on the mining algorithm of miner nodes \ref{MinerALG} in our blockchain. We use Magma \cite{BOSMA1997} to experimentally simulate the miner nodes' mining algorithm to show the efficiency of our proposed PoW model. Tables \ref{table_31}, \ref{table_32}, and \ref{table_33} show the time and memory a node takes to participate in the mining algorithm. The machine environment is a DELL workstation with two Intel Intel(R) Core(TM) i7-9700 CPUs (4 cores, 3.00 GHz per core), 16 GB of memory, and a Windows 10 operating system. We simulated seven sets of systems of random quadratic equations in the finite field $q$ = 2, $q$ = 16, and $q$ = 32. We also used the Gr\(\rm{\ddot o}\)bner basis algorithm, F$_4$, to simulate the consumption of the potential miner nodes. The average time and memory consumption for 100 simulations are shown in the following tables.

\begin{table}[htb]
	\caption{Result of experiments with F$_4$ on random quadratic equations with $q$ = 2}
	\label{table_31}
	\centering
	\begin{tabular}{|c|c|c|}\hline
		\centering
		{\textbf{Parameters}} & {\textbf{Time (s)}} & {\textbf{Memory (MB)}} \\\hline
		\(n=6,m=6\) &0.00    &3.1 \\\hline
		\(n=8,m=8\) &0.031  &3.4 \\\hline
		\(n=10,m=10\)  &0.14 &6.1 \\\hline
		\(n=11,m=11\)  &0.58 &13.3 \\\hline
		\(n=12,m=12\)  &2.84 &41.6 \\\hline
		\(n=13,m=13\)  &13.187 &141.6 \\\hline
		\(n=14,m=14\)  &82.09  &532.0 \\\hline
	\end{tabular}
	\flushleft
\end{table}

\begin{table}[htb]
	\caption{Result of experiments with F$_4$ on random quadratic equations with $q$ = 16}
	\label{table_32}
	\centering
	\begin{tabular}{|c|c|c|}\hline
		\centering
		{\textbf{Parameters}} & {\textbf{Time (s)}} & {\textbf{Memory (MB)}} \\\hline
		\(n=6,m=6\) &0.00    &3.3  \\\hline
		\(n=8,m=8\) &0.031  &3.7 \\\hline
		\(n=10,m=10\)  &0.76 &8.7 \\\hline
		\(n=11,m=11\)  &4.70 &22.0 \\\hline
		\(n=12,m=12\)  &33.75 &75.2 \\\hline
		\(n=13,m=13\)  &241.21 &265.9 \\\hline
		\(n=14,m=14\)  &1854.31  &1009.8 \\\hline
	\end{tabular}
	\flushleft
\end{table}

\begin{table}[htb]
	\caption{Result of experiments with F$_4$ on random quadratic equations with $q$ = 32}
	\label{table_33}
	\centering
	\begin{tabular}{|c|c|c|}\hline
		\centering
		{\textbf{Parameters}} & {\textbf{Time (s)}} & {\textbf{Memory (MB)}} \\\hline
		\(n=6,m=6\) & 0.00    & 3.2  \\\hline
		\(n=8,m=8\) & 0.031  & 3.6 \\\hline
		\(n=10,m=10\)  & 1.25 & 8.8 \\\hline
		\(n=11,m=11\)  & 8.34 & 22.6 \\\hline
		\(n=12,m=12\)  & 63.82 & 77.5 \\\hline
		\(n=13,m=13\)  & 470.81 & 274.6 \\\hline
		\(n=14,m=14\)  & 1048  & Out of memory. \\\hline
	\end{tabular}
	\flushleft
\end{table}

As we can see from the simulation results in these three tables, the work by a miner node contains not only computation capability but also storage capability. We can see that the larger the value of $q$, $n$, and $m$, the larger the time and memory requirement. We provide parameter recommendations based on these phenomena.

In addition, we know that interval block generation time for constructing Bitcoin is approximately 10 min; however, it may be unsuitable and some cryptocurrencies take only 2 min. We fix our interval block generation time to 1 min. We also fix a modern personal computer with a 3 GHz frequency processor, which supplies a computation power of approximately $2^{31}$ operations per second. We are interested in the number of parameters, $n$, of variables we can find with the Gr\(\rm{\ddot o}\)bner basis algorithm, such as F$_4$, in approximately 1 min when we have $m$ equations of degree 2 for different $q$. More precisely, the computer can supply a computation power of approximately $2^{37}$ operations per 1 minute.

Then, in the simulation experiments in Magma, we prefer that the mining process of our PoW algorithm is chosen such that a large amount of memory is not required, that is, typically less than 350 MB \cite{Patarin2020}. Thus, we try to test the parameters using the finite field $q$ from 2 to 32. According to the theoretical complexity of F$_4$, we modify $m$ and $n$ to test and provide our recommend parameter settings. Finally, we obtain the simulation results, as shown in Table \ref{table_4}.

\begin{table*}[ht]
	\centering
	\footnotesize
	\caption{Comparison between TPS on Blockchain coins}
	\label{table_61}
	\begin{tabular}{|c|c|c|c|c|c|c|}
		\hline
		\textbf{Schemes}  & \textbf{Security} & \textbf{PK}  & \textbf{Signature}  & \textbf{Transactions}  \\
		\textbf{(type)}  & \textbf{level} & \textbf{(byte)} & \textbf{(byte)}  & \textbf{(per second)} \\
		\hline
		
		RSA-1024 (ZeroCoin)    &80   &320 &128  &1  \\\hline
		ECDSA-160 (Bitcoin)  &80   &20  &40  &7  \\\hline
		BLISS (Hcash)     &80   &2976  &2720  &0.1  \\\hline
		ID-Rainbow (Our) &80   &8  &46  &24  \\\hline
		Lightweight (Our) &80   &8  &0  &168  \\\hline

	\end{tabular}
	
\end{table*}

\begin{table}[htb]
	\caption{Recommended parameters and result of experiments with F$_4$ on random quadratic equations in approximate one minute}
	\label{table_4}
	\centering
	\begin{tabular}{|c|c|c|}\hline
		\centering
		{\textbf{Parameters}} & {\textbf{Time (s)}} & \textbf{{Memory (MB)}} \\\hline
		\(q=2, m=86, n=28\) &41.22   &314.5  \\\hline
		\(q=4, m=43, n=21\) &86.43   &194.2  \\\hline
		\(q=5, m=40, n=20\) &86.12   &140.7  \\\hline
		\(q=7, m=35, n=19\) &58.65   &112.9 \\\hline
		\(q=8, m=34, n=19\) &65.04   &111.2  \\\hline
		\(q=11, m=30, n=18\) &57.03   &96.0  \\\hline
		\(q=13, m=30, n=18\) &58.39   &97.3  \\\hline
		\(q=16, m=29, n=18\) &58.25   &97.7  \\\hline
		\(q=17, m=27, n=17\) &65.92   &127.2  \\\hline
		\(q=19, m=24, n=16\) &40.25   &93.2  \\\hline
		\(q=23, m=20, n=15\) &35.1   &93.4  \\\hline
		\(q=29, m=16, n=14\) &61.18   &123.6  \\\hline
		\(q=32, m=15, n=13\) &32.95   &51.7  \\\hline
		\(q=32, m=12, n=12\) &64.52   &77.5  \\\hline
	\end{tabular}
	\flushleft
\end{table}

As shown in Table \ref{table_4}, the experiments are conducted for 50 iterations to obtain the average results; we can see that the rate of time requirement and memory requirements are roughly consistent for different parameters. These results show the stability of our PoW algorithms.

\subsection{Comparison on the Transaction Part}

A comparison of the transaction part in our blockchain with the current blockchain (ZeroCoin \cite{Miers2013}, Bitcoin  \cite{Nakamoto2008}, and Hcash \cite{Hcash}) is shown in Table \ref{table_61}. In cryptocurrency blockchain, it  achieves roughly 7 TPS for Bitcoin with 1 MB block and 10 min intervals.  If we introduce the post-quantum signature scheme ``BLISS" \cite{Ducas2013}, the TPS will be reduced to 0.01 TPS. Or, similar to the famous coin Hcash (HC), introduce another post-quantum signature scheme ``BLISS", and the TPS will be reduced to 0.1. 

Then, as seen from Table \ref{table_61}, we have a post-quantum identity-based signature scheme, for example, identity-based Rainbow (ID-Rainbow), with a length of identity of 8 bytes and a signature of 46, the TPS will be 24. Furthermore, once we construct the post-quantum transaction into a lightweight post-quantum segregation witness, the TPS will be improved by seven times.

\section{Conclusion and Future Work}  \label{Sec:Conclusion}

In this paper, we discussed the disadvantages of modern blockchains and their vulnerability to an adversary who has quantum computing power. Then, we proposed a new post-quantum PoW consensus protocol, which can be used to secure the blockchain in place of the existing classical hash-based PoW model.  In addition, we provide a post-quantum transaction mechanism by embedding an identity-based post-quantum signature scheme, which is more suitable for the lightweight transaction implementation in a PQB. This is a secure post-quantum model for blockchain-driven smart cities. Moreover, this work can also help enrich the research on the future blockchain in post-quantum age. In the future, we plan to deploy our construction into a real blockchain.

\section*{Acknowledgment}

We thank the anonymous reviewers' comments and suggestions on our manuscript. This work was partially supported by National Natural Science Foundation of China (Grant No. 61902079  and Grant No. 62002136), the Key Areas Research and Development Program of Guangdong Province (Grant No. 2019B010139002),  and the project of Guangzhou Science and Technology (Grant No. 201902020006 and Grant No. 201902020007).


\section*{References}
\bibliographystyle{model1-num-names}
\bibliography{main}

\end{document}